\documentstyle[epsfig,12pt]{article}

\newcommand{\be}{\begin{equation}}
\newcommand{\ee}{\end{equation}}

\title{
{\bf Rescattering Effects on Intensity Interferometry}}
\author{{J. I. Kapusta and Y. Li} \vspace*{0.1in}\\
 {\it School of Physics and Astronomy, University of Minnesota}\\
 {\it Minneapolis, MN 55455, USA}}

\date{}

\begin{document}

\maketitle

\begin{abstract}
We derive a general formula for the correlation function of two identical particles with the inclusion of multiple elastic scatterings in the medium in which the two particles are produced.  This formula involves the propagator of the particle in the medium.  As illustration of the effect we apply the formula to the special case where the scatterers are static, localized
2-body potentials. In this illustration both $R^2_{\rm side}$ and
$R^2_{\rm out}$ are increased by an amount proportional to the square of the spatial density of scatterers and to the differential cross section.
Specific numbers are used to show the expected magnitude of the rescattering effect on kaon interferometry.
\end{abstract}

\section{Introduction}

The method of two-particle intensity interferometry was developed by Hanbury-Brown and Twiss (HBT) in the 1950's as a means of determining the dimension of distant astronomical objects \cite{HBT}. It was first applied in subatomic physics by Goldhaber {\it et al.} \cite{GGLP} to study the angular correlations between identical pions produced in $p\bar{p}$ annihilations. It was gradually realized that the correlation of identical particles emitted by highly excited nuclei are sensitive not only to the geometry of the system, but also to its
lifetime. Nowadays HBT interferometry is the primary method to probe the space-time structure and the dynamic evolution of the hot matter created in relativistic heavy ion collisions, where a quark gluon plasma is expected to be formed \cite{QGP}. A lot of work has been done both in the theoretical
\cite{Kopylov}-\cite{Pratt} and experimental \cite{STAR01}-\cite{STAR04} sides (for reviews see \cite{Boal}-\cite{WiedemannHeinz}).

HBT measurements are often performed with charged particle pairs, mostly pions. They suffer the long-range Coulomb interaction between themselves and with the remaining source which is created by the primary protons in the colliding nuclei. As a result of that, corrections have to be made to subtract the Coulomb final state interactions before the measured correlations can give useful information about the emitting source \cite{Bowler}-\cite{Sinyukov}.

Additionally, rescattering after production is another effect that may distort the correlation function obtained from experiments. Although rescattering effects are negligible for the applications in astronomy and elementary particle collisions ($ep$, $pp$ and $p\bar{p}$ collisions), it may be significant in heavy ion collisions. In heavy ion collisions, the number of participating nucleons is large (around 400 for central collisions at Relativistic Heavy Ion Collider (RHIC)) and the produced particle multiplicity is in the thousands. The size of the fireball is so big that the produced particles may be scattered a few times in the medium before flying off to the detectors.  The usual interpretation of HBT interferometry is that it measures the distribution of the last scattering (freeze-out distribution). What if the last scattering is very soft? How could that possibly affect the two-particle correlation function for identical particles? This is one of the questions we address.

This paper is organized as follows. In Sec. 2 we derive a general
non-relativistic formula for the correlation function with the inclusion of rescattering in the medium. In Sec. 3 the generalization to relativistically moving fermions and bosons is worked out.  In all cases the resulting correlation function involves the propagator of the particles moving through the medium to the detector.  In Sec. 4 we perform a calculation in one dimension
with the result that rescattering does not change the correlation function.
We also perform a calculation in three dimensions for a static distribution of scattering centers and weak 2-body potentials.  The rescattering effects on the correlation function and on HBT radius parameters are given. The analytical formulas for $R_{\rm out}$ and $R_{\rm side}$ are derived since their combination will give us information on the spatial and temporal distribution of the produced matter. Both $R^2_{\rm out}$ and $R^2_{\rm side}$ are increased by an amount proportional to the square of the spatial density of scatters and to the forward differential cross section. We illustrate these results from by plotting the $R_{\rm out}/R_{\rm side}$ ratio using the experimental parameters from RHIC for kaon interferometry. In Sec. 5 we summarize our results and discuss future applications.

\section{Nonrelativistic Analysis}

In this section we first develop the basic formalism for the two identical particle correlation function.  The particles are produced with a localized wave-packet, are subsequently allowed to interact with the surrounding medium, and are finally observed as momentum eigenstates by a detector.  The limits of chaotic and coherent production are treated explicitly.

\subsection{Basic formalism}

\begin{figure}
\begin{center}
\leavevmode
\hbox{%
\epsfxsize=5.0in
\epsffile{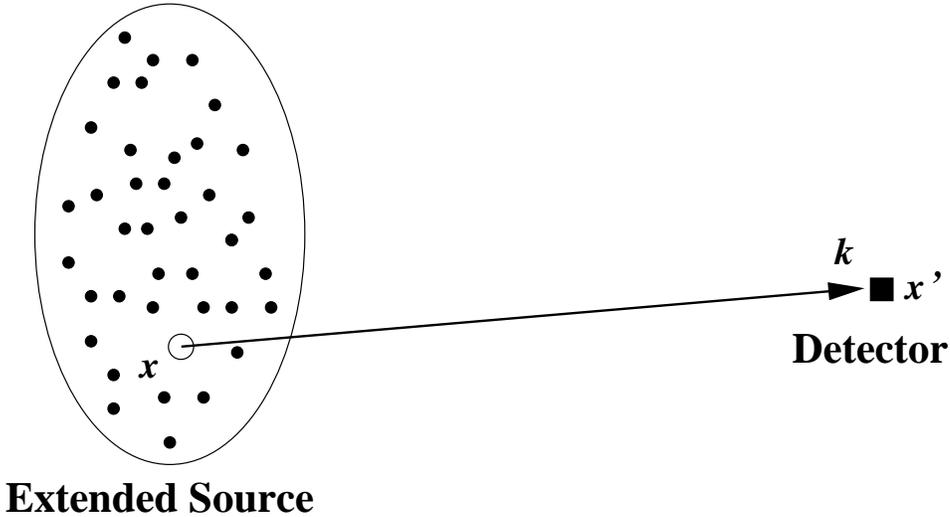}}
\end{center}
\caption{A particle is emitted at a typical source point $x$ of an extended source, and is detected with momentum $k$ at the detection point $x'$.}
\label{oneptcl}
\end{figure}

We first study the detection of a single particle of four-momentum
$k=(k^0,{\bf k})$ at the space-time point $x'=(t',{\bf x}')$. We assume that this particle was produced as a localized wave-packet described by
$\psi_{\rm prod}(x)$.  See Fig. \ref{oneptcl}. The probability amplitude for this to occur is the overlap of the original wavefunction, propagated through time from $t$ to $t'$, with the observed wavefunction
\begin{eqnarray}
\int d^3x d^3x'\psi_{\rm obs}^{\ast}({\bf k};x')G({\bf x}',t';{\bf x},t)\psi_{\rm prod}(x) \, .
\end{eqnarray}
Here $G({\bf x}',t';{\bf x},t)$ is the full nonrelativistic propagator, including all interactions with the surrounding medium, and $\psi_{\rm obs}({\bf k};x')$ is the wavefunction of the particle when it is observed. We assume it to be a momentum eigenstate such that $\psi_{\rm obs}({\bf k};x')={\rm e}^{-i(E_kt' - {\bf k}\cdot {\bf x}')}$. (The absolute normalization drops out of the correlation functions.) The produced wavefunction is a superposition of plane waves centered at ${\bf x}_P$ at time $t_P$ with phase $\chi(x_P)$. That is,
\begin{eqnarray}
\psi_{\rm prod}(x)&=&\int \frac{d^3p}{(2\pi)^3}\varphi({\bf p};x_P){\rm e}^{i \chi(x_P)}{\rm e}^{-i [E_p(t-t_P)-{\bf p} \cdot ({\bf x}-{\bf x}_P)]} \, ,
\end{eqnarray}
where $\varphi({\bf p};x_P)$ is the momentum space wave function. Without loss of generality it can be taken to be real and non-negative. The phase $\chi(x_P)$ can be chaotic, coherent, or partially coherent depending on the nature of the source.  We can write the probability amplitude as
\be
{\cal A}({\bf k};x_P;t',t)={\rm e}^{i \chi(x_P)}{\rm e}^{i E_kt'}\int \frac{d^3p}{(2\pi)^3}\varphi({\bf p};x_P){\rm e}^{i(E_pt_P-{\bf p}\cdot {\bf x}_P)}{\rm e}^{-iE_pt} \widetilde{G}({\bf k},t';{\bf p},t)
\label{nonrelamp}
\ee
where
\be
\widetilde{G}({\bf k},t';{\bf p},t) =
\int d^3x d^3x'G({\bf x}',t';{\bf x},t){\rm e}^{i{\bf p}\cdot{\bf x}}
{\rm e}^{-i{\bf k}\cdot{\bf x}'} \, .
\ee

The single-particle momentum distribution $P({\bf k})$, which is the probability for a particle of momentum $k$ to be produced from the extended source $x$ and arrive at the detection point $x'$, is the absolute square of the total probability amplitude,
\begin{eqnarray}
P({\bf k})&=&\bigg|\sum_{x_P}{\rm e}^{i \chi(x_P)} {\rm e}^{iE_kt'} \int \frac{d^3p}{(2\pi)^3}\varphi({\bf p};x_P) {\rm e}^{i(E_pt_P-{\bf p} \cdot {\bf x}_P)} {\rm e}^{-iE_pt} \widetilde{G}({\bf k},t';{\bf p},t)\bigg|^2 \nonumber \\
& \equiv &\bigg|\sum_{x_P} {\rm e}^{i\chi(x_P)} K({\bf k};x_P;t',t)\bigg|^2 \, .
\label{P}
\end{eqnarray}
Here we define a function
\be
K({\bf k};x_P;t',t) = {\rm e}^{iE_kt'} \int
\frac{d^3p}{(2\pi)^3}\varphi({\bf p};x_P) {\rm e}^{i(E_pt_P-{\bf p} \cdot {\bf x}_P)} {\rm e}^{-iE_pt} \widetilde{G}({\bf k},t';{\bf p},t)
\label{defineK}
\ee
as a folding of the propagator with the momentum-space wave function of the produced wave-packet.

\begin{figure}
\begin{center}
\leavevmode
\hbox{%
\epsfxsize=5.0in
\epsffile{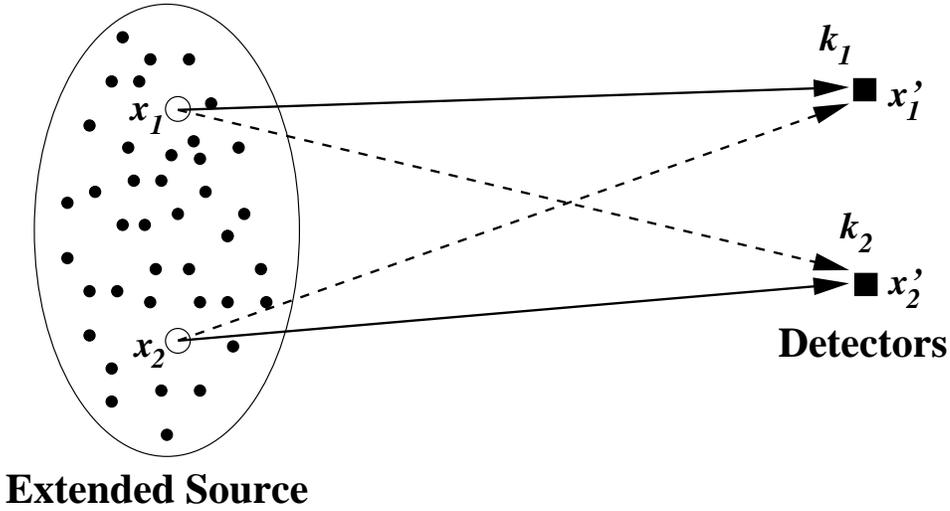}}
\end{center}
\caption{A particle of momentum $k_1$ is detected at $x_1'$ and another identical particle with momentum $k_2$ is detected at $x_2'$. They are emitted from the source points $x_1$ and $x_2$.}
\label{twoptcl}
\end{figure}

Now we turn to the detection of two identical particles. Consider a particle with four-momentum $k_1$ detected at $x_1'$ and another with four-momentum $k_2$ detected at $x_2'$, as shown in Fig. \ref{twoptcl}. The probability amplitude for this occurrence is the product of ${\cal A}({\bf k}_1;x_{1P};t',t)$ and ${\cal A}({\bf k}_2;x_{2P};t',t)$. Because of the indistinguishability of the particles, the probability amplitude must be symmetric with respect to the interchange of two particles if they are bosons or if they are fermions in a spin-singlet state, and antisymmetric if they are fermions in a spin triplet state. The normalized probability amplitude becomes
\begin{displaymath}
\frac{1}{\sqrt{2}}\big\{{\cal A}({\bf k}_1;x_{1P};t',t){\cal A}({\bf k}_2;x_{2P};t',t) \pm {\cal A}({\bf k}_1;x_{2P};t',t){\cal A}({\bf k}_2;x_{1P};t',t)\big\} \, ,
\end{displaymath}
where the plus or minus sign is chosen to make the amplitude symmetric or antisymmetric, respectively.
Therefore, the two-particle momentum distribution $P({\bf k}_1,{\bf k}_2)$ is
\begin{eqnarray}
P({\bf k}_1,{\bf k}_2)&=&\frac{1}{4}\ \bigg|\sum_{x_{1P},x_{2P}}{\rm e}^{i\chi(x_{1P})}{\rm e}^{i\chi(x_{2P})} \Big[K({\bf k}_1;x_{1P};t',t) K({\bf k}_2;x_{2P};t',t)\nonumber \\
& & \pm K({\bf k}_1;x_{2P};t',t)K({\bf k}_2;x_{1P};t',t)\Big]\bigg|^2 \, .
\end{eqnarray}
The {\em intensity correlation function} $C_2({\bf k}_1,{\bf k}_2)$ is defined as the ratio of the probability for the coincidence of ${\bf k}_1$ and ${\bf k}_2$ relative to the probability of observing ${\bf k}_1$ and ${\bf k}_2$ separately,
\begin{eqnarray}
C_2({\bf k}_1,{\bf k}_2)& \equiv &\frac{P({\bf k}_1,{\bf k}_2)}{P({\bf k}_1)P({\bf k}_2)} \, .
\end{eqnarray}

\subsection{Chaotic source}

A chaotic source is one whose phase $\chi(x_P)$ associated with the source point $x_P$ is a random function of the source point coordinate. To take advantage of the randomness of the phases, we expand the right-hand side of Eq. (\ref{P}) into terms independent of $\chi(x_P)$ and terms containing $\chi(x_P)$. We thus obtain
\begin{eqnarray}
P({\bf k})&=&\sum_{x_P}\Big|K({\bf k};x_P;t',t)\Big|^2 +
\sum_{x_P \ne y_P}{\rm e}^{i\chi( x_P)}{\rm e}^{-i\chi(y_P)}
K({\bf k};x_P;t',t)K^{\ast}({\bf k};y_P;t',t) \, .
\end{eqnarray}
The second term in the above equation gives a zero contribution because the large number of terms with slowly varying magnitudes but rapidly fluctuating random phases cancel out one another in the sum. Therefore, the single-particle distribution becomes
\be
P({\bf k})=\sum_{x_P}\Big|K({\bf k};x_P;t',t)\Big|^2
=\int d^4x_P\rho(x_P)\Big|K({\bf k};x_P;t',t)\Big|^2
\ee
for a continuous source, where $\rho(x_P)$ is the density of the source points per unit space-time volume at the point $x_P$.

Similarly, we can rewrite the two-particle distribution function $P({\bf k}_1,{\bf k}_2)$ using the random nature of the chaotic phases. We obtain
\begin{eqnarray}
P({\bf k}_1,{\bf k}_2)&=&\frac{1}{2}\sum_{x_{1P},x_{2P}}\bigg|K({\bf k}_1;x_{1P})K({\bf k}_2;x_{2P}) \pm K({\bf k}_1;x_{2P})K({\bf k}_2;x_{1P})\bigg|^2\nonumber \\
& &+\ \frac{1}{4}\sum_{ {\{x_{1P}x_{2P}\}\ne \{y_{1P}y_{2P}\}}}\Bigg\{{\rm e}^{i(\chi(x_{1P})+ \chi(x_{2P})-\chi(y_{1P})-\chi(y_{2P}))}\Big[K({\bf k}_1;x_{1P})K({\bf k}_2;x_{2P})\nonumber \\
& & \pm \ K({\bf k}_1;x_{2P})K({\bf k}_2;x_{1P})\Big]\Big[K^{\ast}({\bf k}_1;y_{1P})K^{\ast} ({\bf k}_2;y_{2P}) \pm K^{\ast} ({\bf k}_1;y_{2P})K^{\ast} ({\bf k}_2;y_{1P})\Big]\Bigg\}\nonumber \\
&=&P({\bf k}_1)P({\bf k}_2) \pm \bigg|\sum_{x_P}K({\bf k}_1;x_P;t',t)
K^{\ast}({\bf k}_2;x_P;t',t)\bigg|^2 \, .
\end{eqnarray}
Here the only surviving terms are those with $x_{1P}=y_{1P}$ and $x_{2P}=y_{2P}$, or $x_{1P}=y_{2P}$ and $x_{2P}=y_{1P}$. The correlation function $C_2({\bf k}_1,{\bf k}_2)$ becomes
\begin{eqnarray}
C_2({\bf k}_1,{\bf k}_2)&=&1 \pm \bigg|\int d^4x_P \,
{\rm e}^{iq \cdot x_P}\rho_{\rm eff}({\bf k}_1,{\bf k}_2;x_P)\bigg|^2 \, ,
\label{C2}
\end{eqnarray}
where $q=k_1-k_2$ is the difference of the two detected four-momenta. This involves the {\em effective density function} $\rho_{\rm eff}({\bf k}_1,{\bf k}_2;x_P)$, which is
\begin{eqnarray}
\rho_{\rm eff}({\bf k}_1,{\bf k}_2;x_P)& \equiv &\frac{\Big[K({\bf k}_1;x_P;t',t){\rm e}^{-ik_1\cdot x_P}\Big]}{\sqrt{P({\bf k}_1)}}\frac{\Big[K^{\ast}({\bf k}_2;x_P;t',t){\rm e}^{ik_2\cdot x_P}\Big]}{\sqrt{ P({\bf k}_2)}}\rho(x_P) \, .
\label{rhoeffective}
\end{eqnarray}
This is the essential result of our paper.  It includes the effect of rescattering of the particles as they travel through the medium on the intensity correlation function.  Although the derivation was nonrelativistic, the same formula will be shown to apply when the particles move relativistically, with the appropriate change of the propagator.

It is useful to check this formula in the limit of freely propagating particles.  The free nonrelativistic propagator in position space is
\be
G_0({\bf x}',t';{\bf x},t)=\left[\frac{m}{2\pi i(t'-t)}\right]^{\frac{3}{2}}{\rm exp}\left[\frac{im({\bf x}'-{\bf x})^2}{2(t'-t)}\right]\theta(t'-t)
\ee
and its spatial transform is
\be
\widetilde{G}_0({\bf k},t';{\bf p},t) = (2\pi)^3
\delta({\bf k}-{\bf p})\theta(t'-t) {\rm e}^{-iE_k(t'-t)} \, .
\label{freeG}
\ee
This gives the standard result
\be
\rho_{\rm eff}({\bf k}_1,{\bf k}_2;x_P) = \frac{\varphi({\bf k}_1,x_P)
\varphi({\bf k}_2,x_P)}{\sqrt{P({\bf k}_1)P({\bf k}_2)}} \rho(x_P) \, .
\label{free}
\ee
When the produced wave packet in momentum space is independent of production point then $\rho_{\rm eff} = \rho$.

\subsection{Coherent source}

An extended coherent source is described by a production phase $\chi(x_P)$ which is a well-behaved function of the source point coordinate $x_P$. The phases $\chi(x_P)$ and $\chi(y_P)$ at two distinct source points $x_P$ and $y_P$ are related. A simple example of a coherent source is the one in which the production phase $\chi(x_P)$ is a constant. A second example is the case when the production phase $\chi(x_P)$ is a simple function of time, independent of the spatial coordinate.  The two-particle momentum distribution
$P({\bf k}_1,{\bf k}_2)$ for bosons can be written as
\begin{eqnarray}
P({\bf k}_1,{\bf k}_2)&=&\frac{1}{4}\sum_{x_{1P},x_{2P},y_{1P},y_{2P}}\Bigg\{{\rm e}^{i(\chi(x_{1P})+ \chi(x_{2P})-\chi(y_{1P})-\chi(y_{2P}))}\nonumber \\
&\times& \bigg[\ K({\bf k}_1;x_{1P})K({\bf k}_2;x_{2P})K^{\ast}({\bf k}_1;y_{1P})K^{\ast}({\bf k}_2;y_{2P})\nonumber \\
& & +K({\bf k}_1;x_{1P})K({\bf k}_2;x_{2P})K^{\ast}({\bf k}_1;y_{2P})K^{\ast}({\bf k}_2;y_{1P})\nonumber \\
& & +K({\bf k}_1;x_{2P})K({\bf k}_2;x_{1P})K^{\ast}({\bf k}_1;y_{1P})K^{\ast}({\bf k}_2;y_{2P})\nonumber \\
& & +K({\bf k}_1;x_{2P})K({\bf k}_2;x_{1P})K^{\ast}({\bf k}_1;y_{2P})K^{\ast}({\bf k}_2;y_{1P})\bigg]\Bigg\} \, .
\end{eqnarray}
All four terms are essentially the same because all the production points are to be summed over. Obviously this is exactly equal to $P({\bf k}_1)P({\bf k}_2)$. Hence the two-particle correlation function $C_2({\bf k}_1,{\bf k}_2)=1$ for a coherent source. The probability for the detection of one boson is independent of the probability for the detection of another boson. The symmetrization and the factorization of the probability amplitude insure that there is no correlation for a coherent source.

\section{Relativistic Analysis}

In the relativistic domain, the generalization is quite natural. For example, spin-0 bosons with mass $m$ moving under a one-body optical potential $V(x)$ satisfy the Klein-Gordon equation \cite{Gyulassy}
\be
(\partial^{\mu}\partial_{\mu}+m^2+V(x))\phi(x)=0 \, .
\ee
The propagator $i\Delta_V(x',x)\equiv\langle 0|T(\phi(x')\phi(x))|0\rangle$ satisfies the integral equation
\be
\Delta_V(x',x)=\Delta_0(x'-x)+\int d^4y\Delta_0(x'-y)V(y)\Delta_V(y,x)
\ee
with $\Delta_0$ being the free Feynman propagator. Naturally, $\Delta_V$ reduces to $\Delta_0$ in the case $V\rightarrow 0$. The general solution to this integral equation involves a summation of positive and negative frequency solutions, that is, $\phi(x)=\phi^{(+)}(x)+\phi^{(-)}(x)$.  We consider only the positive frequency part which is propagated forward in time by
\be
\phi^{(+)}(x')=-\int d^3x\left[\Delta_V(x',x)\stackrel{\leftrightarrow}{\partial}_0\phi^{(+)}(x)\right]
\ee
where $a\stackrel{\leftrightarrow}{\partial}_0 b\equiv a\left(\partial b/\partial t\right)-\left(\partial a/\partial t\right)b$ \cite{BjorkenBook}.

In analogy with the nonrelativistic case, the probability amplitude that a boson is produced at $x$ and detected at $x'$ is the overlap between the propagated initial wavefunction with the observed wavefunction
\be
{\cal A}=-\int d^3xd^3x'
\phi^*_{\rm obs}(k,x')\left[\Delta_V(x',x)\stackrel{\leftrightarrow}
{\partial}_0\phi_{\rm prod}(x)\right]
\label{relbosonamp}
\ee
where
\begin{eqnarray}
\phi_{\rm prod}(x)&=&\int \frac{d^3p}{\sqrt{2E_p(2\pi)^3}} \,
\varphi({\bf p},x_P) \, {\rm e}^{i\chi(x_P)}{\rm e}^{-ip\cdot(x-x_P)} \, ,\\
\phi_{\rm obs}(k,x')&=&\frac{1}{\sqrt{2E_k(2\pi)^3}} \,
{\rm e}^{-ik\cdot x'} \, .
\end{eqnarray}
The normalization factors do not matter since they will cancel out in the correlation function. The $K$ function defined in Eq. (\ref{defineK}) changes to
\be
K({\bf k};x_P;t',t) =
-\int d^3xd^3x'\frac{d^3p}{(2\pi)^3}
\frac{\varphi({\bf p},x_P)}{\sqrt{4E_kE_p}} \,
{\rm e}^{i{\bf p}\cdot({\bf x}-{\bf x}_P)} \,
{\rm e}^{-i{\bf k} \cdot {\bf x}'} \left[\Delta_V(x',x)\stackrel{\leftrightarrow}{\partial}_0
{\rm e}^{-iE_p(t-t_P)}\right] \, .
\label{relKB}
\ee
The effective density $\rho_{\rm eff}$ and correlation function can be calculated if the explicit form of the propagator is given.

It is easy to evaluate the correlation function for free relativistic bosons simply by using the free propagator $\Delta_0$. One obtains exactly Eq. (\ref{free}); this is left as an exercise for the reader.

Similarly, relativistic spin-$\frac{1}{2}$ fermions with mass $M$ obey the Dirac equation
\be
(i\gamma^{\mu}\partial_\mu-M-V(x))\psi(x)=0 \, .
\ee
Unlike bosons, the intrinsic spin degree of freedom can distinguish the same kind of fermions with different spins. The spins of the observed fermions need to be specified explicitly. The positive frequency solution is now propagated by $S'_F(x',x)$, which is
\be
S'_F(x',x)=S_F(x'-x)-i\int dt\,S_F(x'-x)V(x)\gamma_0
\ee
where $S_F(x'-x)$ is the free Feynman propagator from $x$ to $x'$. The amplitude is
\be
{\cal A}=i\int d^3xd^3x'
\psi^{\ast}_{\rm obs}(k,x',s')S'_F(x',x)\gamma_0\psi_{\rm prod}(x) \, .
\label{relfermionamp}
\ee
The observed wavefunction is
\be
\psi_{\rm obs}(k,x',s')=\frac{1}{(2\pi)^{3/2}}\sqrt{\frac{M}{E_k}} \,
{\rm e}^{-ik\cdot x'}u(k,s')
\ee
where $u(k,s')$ is the spinor which is a positive-energy solution of the Dirac equation with momentum $k^\mu$ and spin $s'$. Moreover
\be
\psi_{\rm prod}(x)=\sum_{s=\pm 1/2}\int \frac{d^3p}{(2\pi)^{3/2}}\sqrt{\frac{M}{E_p}}
b(p,s)u(p,s)\,{\rm e}^{i\chi(x_P)} \, {\rm e}^{-ip\cdot(x-x_P)}
\ee
with $b(p,s)$ being the momentum space wavefunction. The $K$ function changes to
\begin{eqnarray}
K({\bf k};x_P;t',t) &=&
\sum_{s=\pm 1/2}\int d^3x
d^3x'\frac{d^3p}{(2\pi)^3}\sqrt{\frac{M^2}{E_kE_p}}b(p,s)
{\rm e}^{-i{\bf k}\cdot{\bf x}'}{\rm e}^{-ip\cdot(x - x_P)} \nonumber \\
& & \times u^{\dag}(k,s')S'_F(x',x)\gamma_0 u(p,s) \, .
\label{relKF}
\end{eqnarray}
The spin should be taken care of in the antisymmetric two-particle amplitude,
\be
{\cal A}_{\rm two \,\, particles}=
\frac{1}{\sqrt{2}}\left[{\cal A}(k_1,x_{1P},s'_1){\cal A}(k_2,x_{2P},s'_2)-
{\cal A}(k_1,x_{2P},s'_1){\cal A}(k_2,x_{1P},s'_2)\right] \, .
\ee

The correspondence between relativistic and nonrelativistic fermions is the substitution of $i S'_F(x',x)\gamma_0$ for $G(x',x)$.  For free relativistic fermions one obtains
\begin{eqnarray}
\widetilde{G}_{\rm free \,\, fermion} & \equiv &
\int d^3x d^3x' \left[ i S'_F(x',x)\gamma_0 \right]
{\rm e}^{i{\bf p}\cdot {\bf x}} {\rm e}^{-i{\bf k}\cdot {\bf x}'} \nonumber \\
&=& i(2\pi)^2 \delta({\bf k}-{\bf p}) \int dq_0 \,
\frac{(q_0\gamma^0 - {\bf k} \cdot \mbox{\boldmath $\gamma$} + M) \gamma_0}
{q_0^2 - E_k^2 + i\epsilon} {\rm e}^{-iq_0(t'-t)} \, .
\end{eqnarray}
In the nonrelativistic limit this reduces to Eq. (\ref{freeG}).

Comparing with the results in Sec. 2, we find that the difference between nonrelativistic and relativistic formulas is just the change of propagators. The nonrelativistic propagator $G(x',x)$ transforms to
\be
G(x',x) \rightarrow -\Delta_V(x',x)\stackrel{\leftrightarrow}{\partial}_0
\label{GRB}
\ee
for relativistic bosons and
\be
G(x',x) \rightarrow iS'_F(x',x)\gamma_0
\label{GRF}
\ee
for relativistic bosons and fermions, respectively. As expected, the amplitudes in Eq. (\ref{relbosonamp}) and (\ref{relfermionamp}) both reduce to Eq. (\ref{nonrelamp}) in the low-energy limit, as already mentioned.

\section{Rescattering from Weak Potentials}

The usual interpretation of HBT interferometry is that it measures the
space-time distribution of the last scattering, the so-called freeze-out distribution. However, if the scattering is very soft and elastic, we should go backwards in time and take into account multiple scattering effects. Then the boson propagator $G({\bf x}',t';{\bf x},t)$ will become
\begin{displaymath}
G_0({\bf x}',t';{\bf x},t)+G_1({\bf x}',t';{\bf x},t)+
G_2({\bf x}',t';{\bf x},t)+\cdots
\end{displaymath}
where $G_n({\bf x}',t';{\bf x},t)$ is the propagator with $n$ scatterings \cite{BjorkenBook}. In 3-dimensions we have
\begin{eqnarray}
G_0({\bf x}',t';{\bf x},t)&=&\left[\frac{m}{2\pi i(t'-t)}\right]^{\frac{3}{2}}
{\rm exp}\left[\frac{im({\bf x}'-{\bf x})^2}{2(t'-t)}\right]\theta(t'-t) \, , \\
G_1({\bf x}',t';{\bf x},t)&=&-i\int d^3x_sdt_s G_0({\bf x}',t';{\bf x}_s,t_s)V({\bf x}_s,t_s)G_0({\bf x}_s,t_s;{\bf x},t) \, , \\
G_2({\bf x}',t';{\bf x},t)&=&(-i)^2\int d^3x_{1s}dt_{1s}d^3x_{2s}dt_{2s} G_0({\bf x}',t';{\bf x}_{2s},t_{2s})V({\bf x}_{2s},t_{2s}) \nonumber \\
&\times& G_0({\bf x}_{2s},t_{2s};{\bf x}_{1s},t_{1s})V({\bf x}_{1s},t_{1s}) G_0({\bf x}_{1s},t_{1s};{\bf x},t) \, .
\end{eqnarray}
For sake of illustration we take the 2-body potential to be both
time-independent and a function only of the distance between the scattering center, located at ${\bf x}_c$, and the scattering point, located at
${\bf x}_s$.
\begin{eqnarray}
V({\bf x}_s,t_s)=V({\bf x}_s-{\bf x}_c)=\int\frac{d^3q}{(2\pi)^3}{\rm e}^{i{\bf q}\cdot ({\bf x}_s-{\bf x}_c)}\widetilde{v}({\bf q}) \, .
\end{eqnarray}
Here $\widetilde {v}({\bf q})$ is the Fourier transform of the 2-body potential.  For a Gaussian potential $\widetilde {v}({\bf q})=\widetilde {v}({\bf 0}){\rm e}^{-{\bf q}^2r_0^2}$ and for a Yukawa potential $\widetilde {v}({\bf q})=\widetilde {v}({\bf 0})/(1+{\bf q}^2r_0^2)$. A weak potential is characterized by a small value of $\widetilde{v}({\bf 0})$.  We assume that the separation between scattering centers is large compared to the range $r_0$.  We also assume small momentum transfer for which $|{\bf q}|r_0\ll 1$, so that a Gaussian representation is adequate.

\subsection{Rescattering in 1-dimension}

As a first illustration consider the one-dimensional problem.  A particle produced at $x_P=0$ and observed at $x=\infty$ has many possibilities for scattering from localized potentials scattered along the $x$-axis.  The particle can propagate directly to the detector.  Or it may scatter once from a potential localized along the positive $x$-axis, or it may back-scatter from a potential localized along the negative $x$-axis.  The former involves a transmission coefficient, the latter a reflection coefficient.  It may scatter twice: from potentials both along the negative $x$-axis, or from one along the negative and the other along the positive $x$-axis, or from both along the positive $x$-axis.  When the transmission coefficient is much greater than the reflection coefficient (high energy and weak potentials) it is easy to show that
\be
K/K_{\rm free} = 1 -i2\pi m\tilde{v}(0)/k -(2\pi)^2m^2\tilde{v}^2(0)/2k^2
+ \cdot\cdot\cdot = \exp(-i2\pi m\tilde{v}(0)/k) \, .
\ee
Since this is independent of the production point $x_P$ the correlation function is unaffected by rescattering and $\rho_{\rm eff}=\rho$.  This is true even when back-scattering is included.

In one dimension, when the energy is high enough and the individual scattering centers are localized, HBT measures the size of the production source where the wave-packets are produced.  Elastic scattering on localized potentials does not affect this result.  This is consistent with an analysis in \cite{Wong} which considered high energy rescattering in 3-dimensions using the Glauber approach.

\subsection{Rescattering in 3-dimensions}

Next consider rescattering in three dimensions.  We assume for simplicity of illustration that the distribution of the scattering centers is Gaussian,
\begin{eqnarray}
\rho_{\rm scat}({\bf x}_c)=\frac{N_c}{(\sqrt{2\pi}R_c)^3} \,
{\rm e}^{-{\bf x}_c^2/2R_c^2} \, .
\end{eqnarray}
Here $R_c$ and $N_c$ are the radius and the total number of the scattering centers, respectively.  We also assume that the source function is Gaussian and parametrized as
\be
\rho(x)=\frac{1}{4\pi^2R_P^3\tau} \, {\rm e}^{-{\bf x}^2/2R_P^2}
\, {\rm e}^{-t^2/2\tau^2} \, .
\ee
After averaging over all scattering centers, we get
\begin{eqnarray}
K_0({\bf k};x_P;t',t)&=&{\rm e}^{ik\cdot x_P}\varphi({\bf k}) \, ,\\
K_1({\bf k};x_P;t',t)&=&-i\frac{\rm A}{|{\bf k}|}{\rm exp}\Big\{-{\bf x}_{P\perp}^2/2l_c^2\Big\}K_0({\bf k};x_P;t',t) \, ,\\
K_2({\bf k};x_P;t',t)&=&-\frac{\rm A^2}{2|{\bf k}|^2}{\rm exp}\Big\{-{\bf x}_{P\perp}^2/l_c^2\Big\}K_0({\bf k};x_P;t',t) \, .
\end{eqnarray}
Here we define $l_c^2=R_c^2+2r_0^2$ and ${\rm A}=m\widetilde{v}({\bf 0})N_c/2\pi l_c^2$. The ${\bf x}_{P\perp}$ is the projection of ${\bf x}_P$ onto the plane which is perpendicular to the momentum ${\bf k}$. We have also assumed that the momentum transfer is small compared to the detected momentum ${\bf k}$, in accordance with the assumption that the 2-body potential is weak. As can be seen from above,
\begin{eqnarray}
\frac{K({\bf k};x_P;t',t)}{K_0({\bf k};x_P;t',t)}&=&{\rm exp}\left[-i\frac{\rm A}{|{\bf k}|}{\rm exp}\big\{-{\bf x}_{P\perp}^2/2l_c^2\big\}\right] \, .
\end{eqnarray}
The one-particle momentum distribution $P({\bf k})$ is unchanged by the multiple scattering. Therefore, the correlation function takes the form
\begin{eqnarray}
C_2({\bf k}_1,{\bf k}_2)&=&1+\Bigg|\int d^4x\rho(x){\rm e}^{iq\cdot x}{\rm exp}\bigg\{i{\rm A}\bigg(\frac{{\rm e}^{-{\bf x}_{\perp 2}^2/2l_c^2}}{|{\bf k}_2|}-\frac{{\rm e}^{-{\bf x}_{\perp 1}^2/2l_c^2}}{|{\bf k}_1|}\bigg)\bigg\}\Bigg|^2 \, .
\label{full2}
\end{eqnarray}

For the two-particle momentum distribution $P({\bf k}_1,{\bf k}_2)$, we are interested in two special cases: ${\bf k}_1 \parallel {\bf k}_2$ and $|{\bf k}_1|=|{\bf k}_2|$, because they will provide information about $R_{\rm out}$ and $R_{\rm side}$, respectively. Define pair momenta ${\bf k}=({\bf k}_1+{\bf k}_2)/2$ and ${\bf q}={\bf k}_1-{\bf k}_2$. If ${\bf k}_1$ is parallel to ${\bf k}_2$, we have ${\bf x}_{\perp 1}={\bf x}_{\perp 2}$ and ${\bf k}\parallel{\bf q}$. Then the correlation function becomes
\begin{eqnarray}
C_2({\bf k}_1,{\bf k}_2)&=& 1+{\rm e}^{-(q_0^2\tau^2+{\bf q}^2R_P^2)}\bigg\{1-\frac{{\rm A}^2 R_P^4l_c^2}{(l_c^2+R_P^2)^2(l_c^2+2R_P^2)}\bigg(\frac{1}{|{\bf k}_1|}-\frac{1}{|{\bf k}_2|}\bigg)^2\bigg\} \, .
\end{eqnarray}
For not too small total momentum $|{\bf k}|$ the source size can be determined from the curvature of the correlation function at $|{\bf q}|=0$ \cite{Pratt}. We get
\begin{eqnarray}
R_{\rm out}^2&=&R_P^2+\frac{{\bf k}^2\tau^2}{m^2}+\frac{{\rm A}^2 R_P^4l_c^2}{(l_c^2+R_P^2)^2(l_c^2+2R_P^2)}\frac{1}{|{\bf k}|^4} \, .
\end{eqnarray}
If ${\bf k}_1$ and ${\bf k}_2$ have the same magnitude, ${\bf k}$ is perpendicular to ${\bf q}$. After the integration in Eq. (\ref{full2}), we obtain
\begin{eqnarray}
R_{\rm side}^2&=&R_P^2+\frac{{\rm A}^2 R_P^4}{(l_c^2+2R_P^2)^2}\frac{1}{|{\bf k}|^4} \, .
\end{eqnarray}
Rescattering increases both effective radii compared to the primordial source radius, as should be expected, albeit by differing amounts.

\subsection{Numerical example: kaons}

As a numerical example, consider kaons scattering from slowly moving nucleons. The forward differential cross section is taken to be 1 mb/sr. The central density of nucleons is taken to be 0.15 nucleons/fm$^3$ and the
radii are $R_P=R_c=5$ fm. The duration of emission is taken to be $\tau = 2$ fm/c.

\begin{figure}[!htbp]
 \centering
 \includegraphics[width=4in]{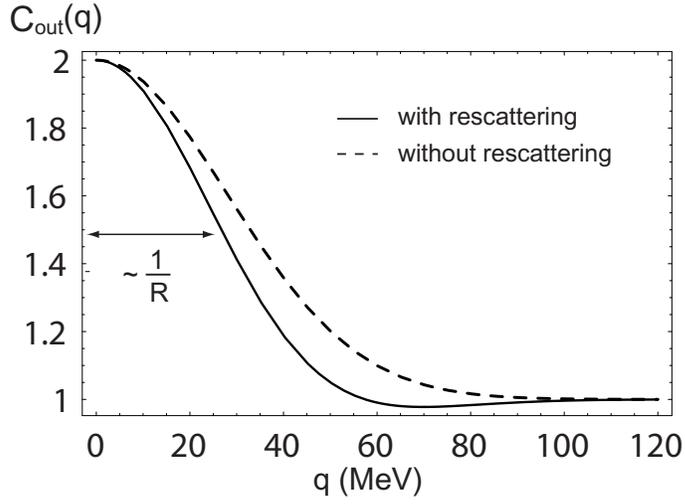}
 \caption{The outward correlation function $C_{\rm out}(q)$ as a function of relative momentum q with the
 parameters specified in the text.}
 \label{fig:Cout}

\end{figure}

\begin{figure}[!htbp]
 \centering
 \includegraphics[width=4in]{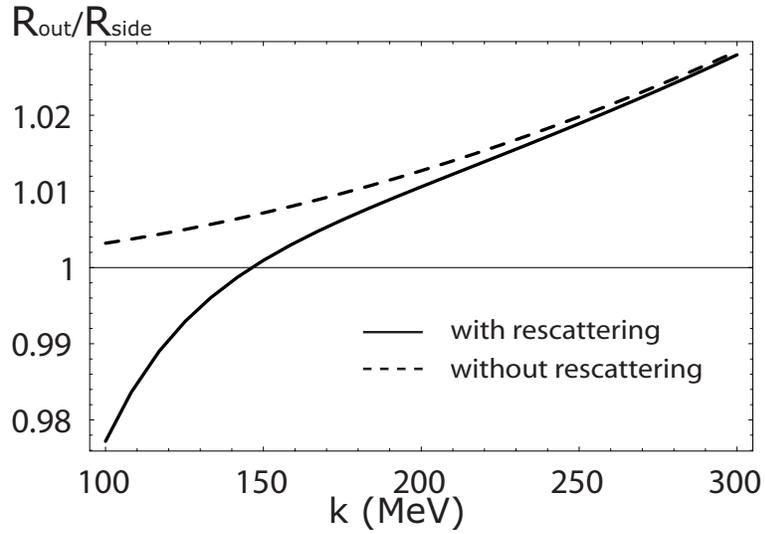}
 \caption{The ratio of outwards to sidewards radii that would be measured for kaons.}
 \label{fig:RoutRside}
\end{figure}

The outward correlation function is shown in Fig. \ref{fig:Cout}. A similar plot can be drawn for the sideward correlation function. Since the HBT radii are inversely proportional to the width of the correlation function, rescattering in the medium tends to increase both HBT radii. The resulting value of
$R_{\rm out}/R_{\rm side}$ is shown in Fig. \ref{fig:RoutRside}. We see that rescattering tends to reduce the $R_{\rm out}/R_{\rm side}$ ratio, especially for low-energy particles which spend a longer time in the medium. The
ratio is less than 1 for $k < 150$ MeV and greater than 1 for $k > 150$ MeV. The ratio is a monotonically increasing function of $k$. The derivation given in Sec. 2 is quite general, but this numerical example does not
include any collective flow of either the produced particles or of the scattering centers (nucleons), which would tend to make the individual radii decrease with increasing momentum. As a result of that, it is not appropriate to compare Fig. \ref{fig:RoutRside} with RHIC experimental data.

\section{Conclusions}

In this paper we have derived explicit formulas for the Hanbury-Brown Twiss intensity interferometry correlation function when the particles undergo rescattering in the medium in which they are produced.  The primary result is encapsulated in Eq. (\ref{C2}) (for chaotic sources).  It is essentially expressed in terms of the Fourier transform of an effective source density given by Eq. (\ref{rhoeffective}).  The latter involves a folding of the propagator of the particle in the medium with the momentum-space wave function.  For nonrelativistic particles one uses Eq. (\ref{defineK}), while for relativistic bosons one uses Eq. (\ref{relKB}) and for relativistic fermions one uses Eq. (\ref{relKF}).  The correspondence between the nonrelativistic and relativistic versions of the propagators is given by Eqs. (\ref{GRB}) and (\ref{GRF}).  It should be emphasized that, in general, these propagators are time-dependent since the medium is most likely expanding and cooling.

For examples, we studied the rescattering from weak potentials in the nonrelativistic limit.  In one dimension there was no effect on the two-particle correlation function.  In three dimensions, we found that both $R^2_{\rm side}$ and $R^2_{\rm out}$ are increased by an amount proportional to the square of the spatial density of scatterers and to the differential cross section.  Specific numbers were used to show the expected magnitude of the rescattering effect on kaon interferometry.  In particular, the $R_{\rm out}/R_{\rm side}$ ratio increases with transverse momentum from below unity to above unity.

This theory now must be applied to real experiments and data on heavy ion collisions.  The general approach is applicable from low energy fragmentation reactions to the highest energy collisions where quark-gluon plasma formation is expected.  Complications such as moving scattering centers, long-lived resonances, flow, partial coherence, and so on must be taken into account for detailed comparison to experimental data.  Such work is in progress.

\section*{Acknowledgements}

This work was supported by the US Department of Energy under grant
DE-FG02-87ER40328.

\end{document}